\documentclass[10pt,conference]{IEEEtran}

\usepackage{amssymb}
\usepackage{pifont}
\usepackage{graphicx}
\usepackage[dvipsnames]{xcolor}
\usepackage{soul}
\usepackage{cite}
\usepackage{amsmath,amssymb,amsfonts}
\usepackage{listings}
\usepackage{algorithmic}
\usepackage{graphicx}
\usepackage{subcaption}
\usepackage{textcomp}
\usepackage{balance}
\usepackage[colorlinks=true,urlcolor=blue,citecolor=black,linkcolor=black]{hyperref}
\usepackage[utf8]{inputenc}
\usepackage[T1]{fontenc}
\usepackage{url}
\usepackage{wrapfig}
\usepackage{caption}
\usepackage{algorithm}
\usepackage{hhline}
\usepackage{fixltx2e}
\usepackage{tabularx}
\usepackage{booktabs}
\usepackage{colortbl}
\usepackage{multirow}
\usepackage{enumitem}
\usepackage{flushend} 
\usepackage{orcidlink}

\usepackage[scaled=0.86]{helvet}

\usepackage{framed}

\definecolor{light-gray}{gray}{0.9}
\usepackage[framemethod=TikZ]{mdframed}

\mdfdefinestyle{MyFrame}{%
    linecolor=black,
    outerlinewidth=0.15pt,
    roundcorner=3pt,
    innertopmargin=2pt,
    innerbottommargin=2pt,
    innerrightmargin=4pt,
    innerleftmargin=4pt,
    backgroundcolor=light-gray}
    
\mdfdefinestyle{MyFrameSimple}{%
    linecolor=black,
    outerlinewidth=0.15pt,
    roundcorner=3pt,
    innertopmargin=2pt,
    innerbottommargin=2pt,
    innerrightmargin=4pt,
    innerleftmargin=4pt}

\renewcommand{\checkmark}{\ding{51}}
\newcommand{\xmark}{\ding{55}}

\ifCLASSINFOpdf
\else
\fi

\begin{document}
\title{Requirements-driven Slicing of Simulink Models Using LLMs}

\author{\IEEEauthorblockN{Dipeeka Luitel\orcidlink{0009-0004-5306-8786}, Shiva Nejati\orcidlink{0000-0002-0281-8231}, Mehrdad Sabetzadeh\orcidlink{0000-0002-4711-8319},}
\IEEEauthorblockA{School of Electrical Engineering and Computer Science\\
University of Ottawa, Canada\\
\{dipeeka.luitel, snejati, m.sabetzadeh\}@uottawa.ca}
}

\maketitle
\pagenumbering{arabic}
\thispagestyle{plain}
\pagestyle{plain} 

\begin{abstract}
Model slicing is a useful technique for identifying a subset of a larger model that is relevant to fulfilling a given requirement. Notable applications of slicing include reducing inspection effort when checking design adequacy to meet requirements of interest and when conducting change impact analysis.  In this paper, we present a method based on large language models (LLMs) for extracting model slices from graphical Simulink models. Our approach converts a Simulink model into a textual representation, uses an LLM to identify the necessary Simulink blocks for satisfying a specific requirement, and constructs a sound model slice that incorporates the blocks identified by the LLM. We explore how different levels of granularity (verbosity) in transforming Simulink models into textual representations, as well as the strategy used to prompt the LLM, impact the accuracy of the generated slices.  Our preliminary findings suggest that prompts created by textual representations that retain the syntax and semantics of Simulink blocks while omitting visual rendering information of Simulink models yield the most accurate slices. Furthermore, the chain-of-thought and zero-shot prompting strategies result in the largest number of accurate model slices produced by our approach.


\end{abstract}

\IEEEpeerreviewmaketitle

\section{Introduction}\label{sec:introduction}
Models have long served as key tools for capturing safety-critical software designs within cyber-physical system (CPS) domains, such as avionics, railways, maritime, and automotive. Certification, which is often a must in these domains, involves meeting strict safety requirements through a detailed review of design models by an assessing body. The certification process requires that assessors meticulously scrutinize the models to identify design elements related to safety-critical requirements -- a task that is challenging and susceptible to errors. This complexity arises because assessors, often external entities, must navigate extensive design models to identify and evaluate the relevant safety aspects, ensuring the design's compliance with safety requirements.

Simulink~\cite{simulink}, a block-based modelling language, is a prominent tool for modelling CPS. Simulink facilitates safety-critical analysis by providing a visual interface for building and simulating dynamic systems. Ensuring that a Simulink model meets safety requirements is time-consuming, as it involves examining the entire model to ensure the preservation of a requirement, which might only pertain to a small subset of the model. This highlights the need for methods such as model slicing~\cite{tip1994}, which enables engineers to automatically extract the safety-related slices of Simulink models. Model slicing breaks down a given model into smaller, manageable pieces, each associated with specific requirements. This targeted approach reduces the complexity involved in understanding and analyzing an entire model, thus making it easier to validate whether a system meets its safety-related requirements.

Model slicing has been applied to a variety of artifacts including requirements models~\cite{binalialhag2019}, behavioural models~\cite{kan2017, taentzer2018}, and other development artifacts~\cite{rahim2021, wang2020}. Model slicing has various applications including change impact analysis~\cite{nejati2016sysml, hajri2018, arora2015nlp} and model verification~\cite{alenazi2020, nejati2011sysml}, among others. Automated slicing techniques often rely on manually created links to establish traceability~\cite{alenazi2020}. Creating these trace links can be time-consuming and error-prone. The emergence of Large Language Models (LLMs) provides the prospect of defining slicing as an end-to-end activity without relying on manual traceability. While machine learning has already been used to automate the creation of trace links (e.g., \cite{zhao2017embeddings, guo2017deeplearning}), the potential of LLMs~\cite{chatgpt2023} to improve the automation of trace links, which could then be used for creating slices or to directly obtain model slices, remains to be explored.

In this paper, we propose using LLMs to automatically generate requirements-based slices for Simulink models, primarily to facilitate design inspections during certification. Given a Simulink model and a natural-language requirements statement, we employ LLMs to extract a slice of the model that satisfies the requirement without the need for predefined trace links. While Simulink models are graphical, they can be converted into textual representations that are amenable to processing by LLMs. In essence, what we propose is a \emph{``both requirements and models as text''} method of slicing, enabling the analysis of natural-language requirements and models side-by-side and in a unified representation.

\begin{figure*}[t]
  \centering
    \includegraphics[width=\textwidth]{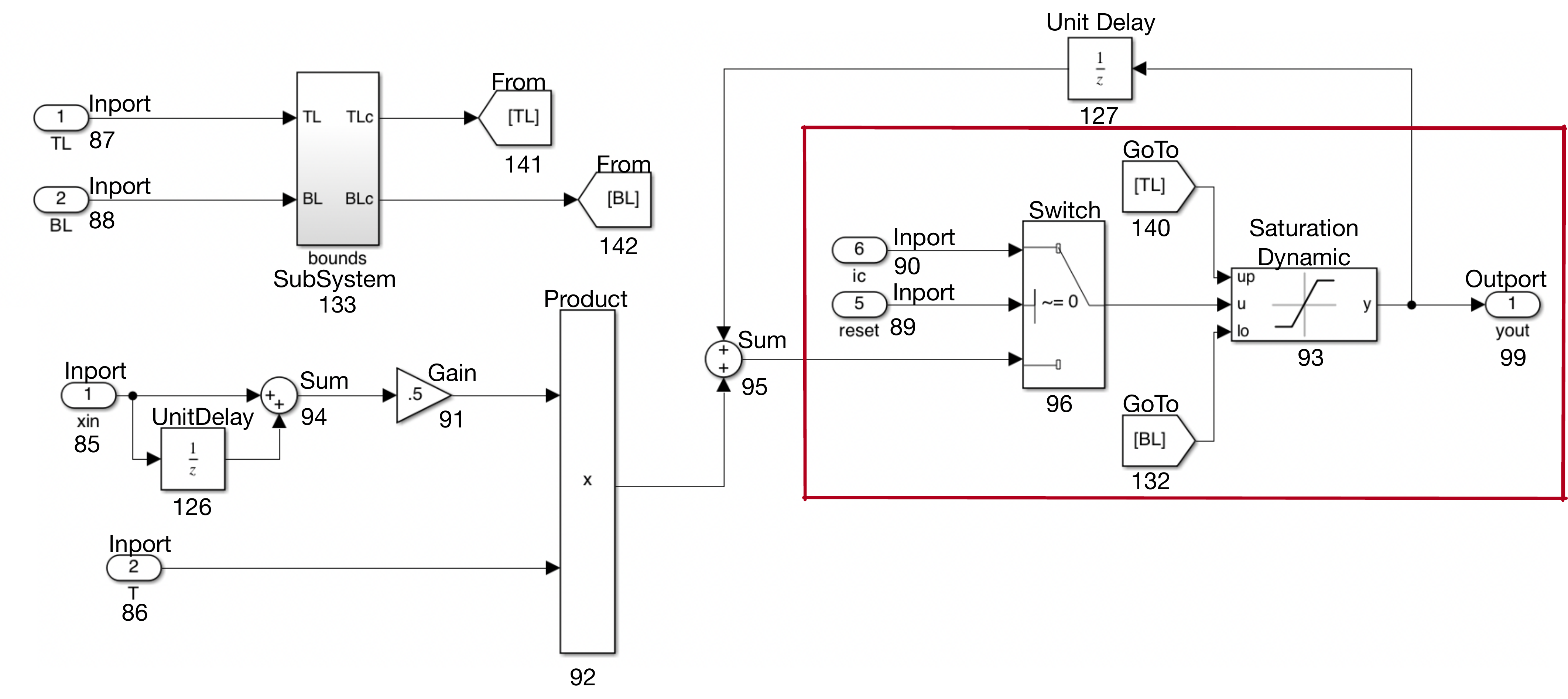}
    \caption{The Tustin integrator model from Lockheed Martin's Simulink benchmark~\cite{mavridou2020lockheed}. The area outlined in red identifies the blocks that contribute to the satisfaction of requirement $R$: ``When $\mathit{reset}$ is True and the Initial Condition ($\mathit{ic}$) is within the Top and Bottom Limits ($\mathit{BL} \leq \mathit{ic} \leq \mathit{TL}$), the Output ($\mathit{yout}$) should match the Initial Condition ($\mathit{ic}$)''. The block type is indicated by the label above it, while the number beneath each block represents its SID.}\label{fig:tustin}
\end{figure*}

Our slicing method has three phases: (1) transforming a given Simulink model into a textual representation, (2) employing an LLM to identify block IDs of the Simulink blocks necessary for fulfilling a given (natural-language) requirement, and (3) generating a model slice by
systematically filtering blocks deemed non-essential by the LLM
for fulfilling the requirement in question. The level of granularity
(verbosity) in the transformation to textual representation, as well
as the strategy used to prompt the LLM for results, can influence
the quality of the generated slice. We validate the accuracy of
slices generated by our approach by executing them and comparing their ability to satisfy requirements with that of the original
model. Our preliminary findings suggest that retaining either
insufficient or excessive information in the textual representation
reduces slicing quality, thus favouring the use of a middle ground
between the two extremes for transformation granularity. Specifically, prompts created using medium-granularity textual representations, which preserve the syntax and semantics of Simulink blocks while excluding visual rendering information, generate the most accurate slices. Moreover, when combined with either the chain-of-thought or zero-shot prompting strategies, the medium-granularity transformations result in the highest number of accurately produced model slices by our approach.

The rest of the paper is structured as follows: We present a running example to motivate our work in Section~II. Section~III provides background information. Section~IV discusses related work. Section~V presents our requirements-driven approach for model slicing. Section~VI reports on our experimental design and evaluation. Section~VII summarizes our work and provides a brief discussion for future research. 
\section{Motivating Example} \label{sec:motivatingexample}
We motivate our work using the Simulink model shown in Fig.~\ref{fig:tustin}, which we refer to as \emph{Tustin}. The figure also shows the Simulink ID (SID) for each block in the Tustin model. The Tustin model, also known as the bilinear transform method, improves the calculation of integrals over time by converting continuous-time systems into discrete-time systems. 
The Tustin model is part of Lockheed Martin's public-domain benchmark of Simulink models~\cite{mavridou2020lockheed} for the cyber-physical system (CPS) industry. The benchmark is comprised of models that exemplify various CPS behaviours in aerospace and defense. Tustin has five functional requirements expressed in natural language that the model is expected to satisfy. However, not all blocks in the Tustin model are relevant to all of these five requirements. Indeed, each of the five requirements can be traced to a subset of the blocks in Tustin. Model slicing isolates specific, relevant sections of a model based on a given requirement. This technique is especially beneficial for focusing on certain aspects of the model, thereby simplifying inspection and verification processes.

Consider requirement $R$: ``When $\mathit{reset}$ is True and the Initial Condition ($\mathit{ic}$) is within the Top and Bottom Limits ($\mathit{BL} \leq \mathit{ic} \leq \mathit{TL}$), the Output ($\mathit{yout}$) should match the Initial Condition ($\mathit{ic}$)''. The blocks fulfilling this requirement are outlined in red in Fig.~\ref{fig:tustin}. These blocks represent a \emph{slice} of the Tustin model implementing the function of $R$ and ensuring its satisfaction. Specifically, the model slice for requirement $R$ includes, among other blocks, a switch block, indicated by block ID 96 in Fig.~\ref{fig:tustin}, and a saturation block, indicated by block ID 93. The switch block ensures that, when $\mathit{reset}$  is True,  the initial condition $\mathit{ic}$  is selected and passed to the saturation block which, subsequently, ensures that the output, $\mathit{yout}$,  is bounded between the  the upper and lower saturation values, i.e., $\mathit{TL}$ and $\mathit{BL}$ respectively. Hence, the the combination of the switch block (ID=96) and the saturation block (ID=93) ensures the satisfaction of $R$.

The goal of requirements-based slicing is to identify a subset of blocks in a Simulink model that are necessary to satisfy a given requirement of that model. Requirements-based slicing for Simulink models involves a considerable volume of textual information, including the requirements statements themselves, the explicit labels and block names provided by modellers, and the graphical representation that can be converted into (textual) markup in a systematic way. In this paper,  we propose an approach for  automatically generating requirements-based  slices for Simulink models using  LLMs. To do so, we convert Simulink models into text and provide systematic prompts for LLMs so that they can identify Simulink blocks pertinent to a given requirement.

\begin{figure}[t]
\centering
\includegraphics[width=\linewidth]{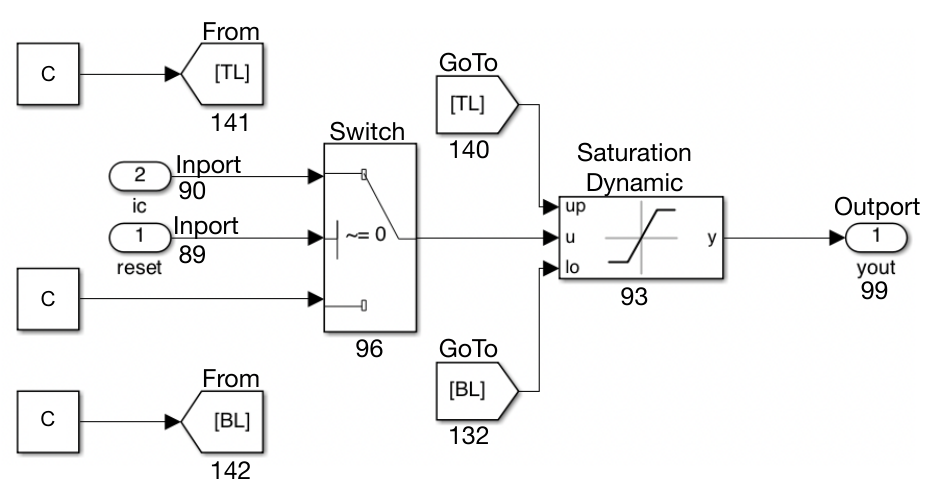}
\caption{A slice of the Tustin model (Fig.~\ref{fig:tustin}) for requirement $R$, obtained using LLM-based slicing. The label above each block specifies its type, and the number below each block denotes its SID.}\label{fig:model_slice}
\vspace*{-1.5em}
\end{figure}

We then construct a model slice from the original model, keeping the identified blocks. For example, the model slice of Tustin computed by our approach for our example requirement, $R$, is shown in Fig.~\ref{fig:model_slice}.

\section{Background}\label{sec:background}
\subsection{Large Language Models (LLMs)}\label{subsec:llm}
Large Language Models (LLMs) are statistical models trained to predict and generate coherent, contextually relevant text. These models are broadly categorized into \emph{generative} models, such as GPT (Generative Pre-training Transformer)~\cite{radford2018improving}, which excel in text production, and \emph{discriminative} models, such as BERT (Bidirectional Encoder Representations from Transformers)~\cite{devlin2018bert}, which are optimized for classification and analysis tasks. LLMs are typically pre-trained on large volumes of textual data to learn contextual information, language regularities, and syntactic and semantic relationships. Our model slicing approach assumes a generative LLM. Specifically, we use ChatGPT~\cite{openai2020gpt} over GPT-4.0-classic for experimentation, noting that we do not require real-time data retrieval from the response generation capabilities of GPT-4.0.

\subsection{Prompt-based Learning} \label{subsec:prompt}
Generative AI models, including generative LLMs, commonly use \emph{prompts} as input instructions to guide the model, enabling users to specify the desired nature or context of the generated output~\cite{hariri2023unlocking}. Various prompting strategies, such as zero-shot, few-shot, chain-of-thought, and tree-of-thought prompting, exist to enhance output quality~\cite{Liu2023pre}. In this paper, we examine zero-shot, N-shot, and chain-of-thought learning for model slicing, investigating whether additional examples in the input prompt would progressively enhance the quality of an LLM's response. In zero-shot learning, the LLM is provided only with task instructions without any examples. This method relies on an LLM's existing extensive training to formulate its responses. For N-shot learning, the prompt includes \emph{n} number of examples, using demonstrative teaching to train the LLM's response. The chain-of-thought approach incorporates a series of logical reasoning steps in the examples, clearly outlining the expected reasoning process to the LLM.

\subsection{Simulink}
The models subjected to requirements-based slicing in this paper are Simulink models. Simulink is a MATLAB-based graphical modeling environment used for  simulating and analyzing dynamic systems. Simulink has numerous applications in disciplines such as control systems, signal processing, communications, and software engineering. In software engineering, the most notable application of Simulink is for modelling cyber-physical systems (CPS), which integrate the digital and physical worlds through interconnected computational networks. Simulink employs a visual block diagram approach wherein mathematical functions are encapsulated in blocks interconnected to capture data flow. These models are organized into a hierarchical framework of subsystems, each characterized by designated inputs and outputs, encompassing Simulink blocks and potentially additional subsystems. Configurable parameters are essential for model adaptation to specific hardware implementations. Simulink distinguishes between stateless blocks, whose outputs depend exclusively on present inputs, and stateful blocks, whose outputs are influenced by both current inputs and their internal states, with further classification into discrete and continuous types based on state-transition dynamics.

\section{Related Work} \label{sec:related}
We position our research within the literature, covering existing approaches to (1) interpreting models as textual representations, (2) model slicing, and (3) change impact analysis.

Griebl et al.~\cite{griebl2023scratch} investigate using language models, such as n-gram models, for code completion and bug detection for the block-based programming language, SCRATCH. To do so, they treat graphical SCRATCH programs as textual representations. We too view block models in a textual format; however, our work is distinguished from that of Griebl et al. in the following ways: (1) we focus on Simulink block diagrams, (2) we have different analysis goals, and (3) our analysis task is driven by requirements, whereas Griebl et al.'s approach focuses on n-gram models, aiming to predict the most likely sequence of blocks for code completion and identify the least likely sequence of blocks for bug detection.

Alenazi et al.~\cite{alenazi2020} use a mutation-driven approach for tracing requirements in state-based design models. The generated model mutants are verified through model checking to establish traceability by identifying co-occurrence patterns in killed versus survived mutants. Nejati et al.~\cite{nejati2012sysml, nejati2011sysml} introduce SafeSlice, a tool that implements a SysML-based framework for managing traceability and design slicing. SafeSlice aims to address the traceability gap between safety requirements and software design, which is essential for software safety certification. The tool includes a traceability information model, a methodology for establishing traceability, and an algorithm for extracting design slices relevant to safety requirements. These two works use a manual approach for establishing tractability. In contrast, our work uses LLM-based automation for slicing and does not involve the creation of explicit traceability links as an intermediate step.

Change impact analysis and model slicing are similar in that both tasks analyze the relevance of elements to a particular artifact or event. Nejati et al.~\cite{nejati2016sysml} present an approach for automated change impact analysis between SysML models of requirements and design. In this work, the authors introduce a method that uses a combination of static slicing algorithms and NLP to scope the impact of requirement changes on system design elements. Hajri et al.~\cite{hajri2018} also perform change impact analysis, in the context of evolving configuration decisions within use case models. They use NLP for identifying the effects of decision changes on product-line use case diagrams to automatically regenerate product-specific use case models. Arora et al.~\cite{arora2015nlp} use an NLP-based approach for performing change impact analysis on natural-language requirements. They automatically identify the phrasal structure of requirements and quantify the likelihood of each statement being impacted by a change.
Goknil et al.~\cite{goknil2014} present a tool which uses a metamodeling approach for change impact analysis in software requirements. They model requirements and their interrelations for automated inference and consistency checking using formal semantics to describe requirements relations. Unlike the papers mentioned above, which employ a broad strategy of system-wide tracing to evaluate change impacts, we focus on tracing requirements to model blocks that  contribute to the satisfaction of those requirements. Furthermore, our research is specifically targeted at Simulink models, which, to our knowledge, complement prior work on both model slicing and change impact analysis.
\section{Approach}\label{sec:appraoch}
\begin{figure}[t]
\centering
\includegraphics[width=\linewidth]{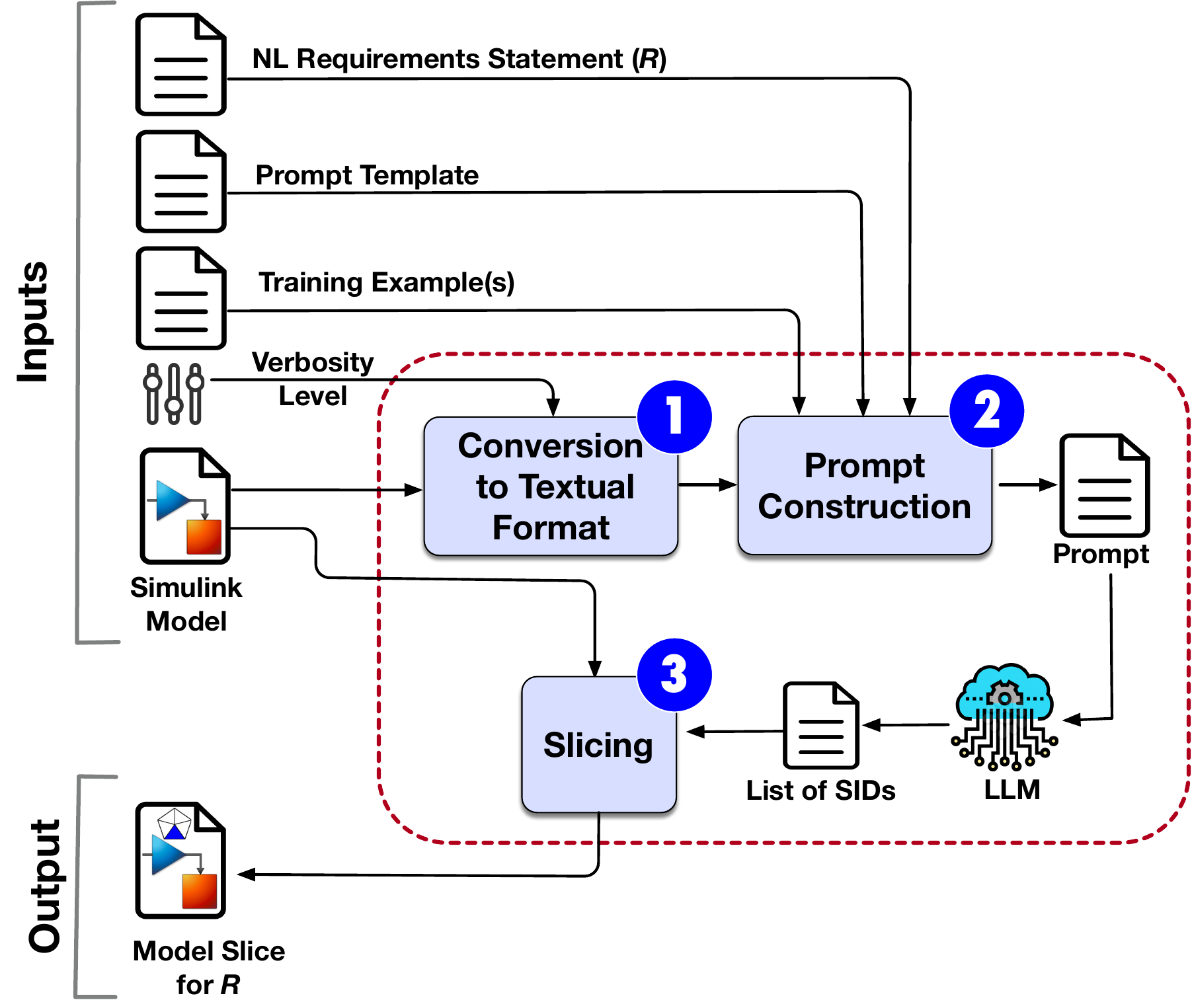}
\caption{An overview of our LLM-based approach: The approach has five inputs and produces, as output, a slice of the input Simulink model for requirement $R$.}\label{fig:approach}
\end{figure}

Figure~\ref{fig:approach} presents an overview of our approach. The approach has five inputs: (1)~a Simulink model, (2)~a natural-language requirements statement ($R$) for which we would like to compute a model slice, (3)~a prompt template for extracting Simulink blocks relevant to $R$, (4)~training example(s) used for prompt-based learning, and (5)~a verbosity level that determines the granularity of the transformation from the Simulink model into its textual representation. The approach output is a slice of the input Simulink model derived for $R$.

\subsection{Conversion To Textual Format}\label{subsec:conversion}
The first step of our approach is to convert the input Simulink model, which is  graphical, into a textual format. Simulink models already come with an XML-based textual representation that is supported by MATLAB. The XML schema includes a variety of attributes, such as block name, type, Simulink ID (SID), block properties, and block coordinates for visual rendering. These attributes collectively cover the semantics, syntax, and graphical information related to a Simulink model. For instance, while block type conveys critical semantic information, its coordinates are irrelevant to its function. In our approach, we allow the user to choose the verbosity level, providing flexibility in determining the degree of detail in the textual representation of a Simulink model. The levels of verbosity reflect the granularity of transformation (i.e., extent of model abstraction when converting a Simulink model into text). LLMs are more likely to hallucinate when given lengthy and irrelevant input prompts. This is less of an issue with smaller models, but becomes significant with large Simulink models. These larger models may require greater abstraction (i.e., using a low verbosity level) to ensure only essential block content is preserved for preventing hallucinations. This enables us to experiment with different verbosity levels to identify which level is best suited for LLM-based slicing. Being able to adjust the verbosity level is also critical for ensuring that the textual representation of large Simulink models will fit within the underlying LLM's token limit, the maximum tokens the LLM can process in one input sequence.

\begin{figure}
\centering\includegraphics[width=.955\linewidth]{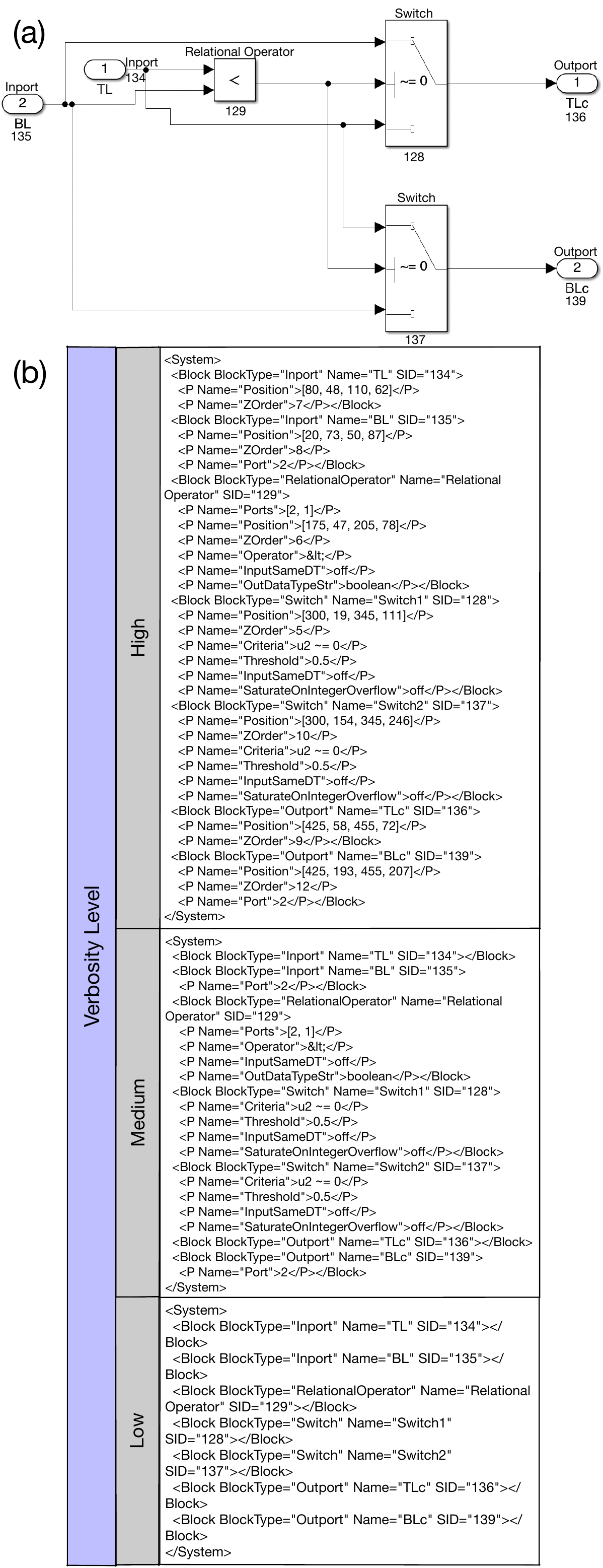}
    \caption{Converting Simulink models to text: (a)~An excerpt of the Tustin model  in Fig.~\ref{fig:tustin}, and (b) textual representations of the model excerpt for the high, medium and low verbosity levels.}
    \label{fig:simulink_example}
\end{figure}

Figure~\ref{fig:simulink_example} illustrates how a fragment of the Tustin model in Fig.~\ref{fig:tustin} is converted into text for three different verbosity levels considered in our work: high, medium, and low. Specifically, the Simulink model fragment in Fig.~\ref{fig:simulink_example}(a) includes two Inport blocks, two Switch blocks, one RelationalOperator block, and two Outport blocks. As shown in Fig.~\ref{fig:simulink_example}(b), the textual representation for the high verbosity level includes all block attributes as well as the attributes related to the visual rendering of the blocks.  At the medium level, we retain block content while omitting visual rendering details. For example, at the medium verbosity level, the text for the Switch blocks omits the coordinates of the blocks' positions within the model but retains all the Switch blocks' attributes related to their semantics.  At the low level, we focus only on the most essential block content: block name, type, and SID. As illustrated in Fig.~\ref{fig:simulink_example}(b), at the low verbosity level, the text capturing the model fragment in Fig.~\ref{fig:simulink_example}(a) is reduced to a basic outline. The textual representation obtained from the input Simulink model is used for \hbox{prompt construction, described next.}

\subsection{Prompt Construction}
The second step of our approach is to construct a prompt for generating a model slice.
Our approach supports three prompting strategies, N-shot, chain-of-thought and zero-shot, discussed in Section~\ref{subsec:prompt}. Among these, zero-shot learning requires no training examples, while N-shot and chain-of-thought strategies need one or more training examples to teach an LLM how to create model slices. To support these two strategies, our approach accepts as input instances of requirements-based slices for Simulink models used only for training  -- different from the model for \hbox{which we aim to compute a slice.}

Figure~\ref{prompt:example} shows our prompt template, which is comprised of six segments: three are fixed, and the remaining three have blue-highlighted placeholders that must be replaced with specific text to transform the template into a complete prompt for generating model slices. The first segment is Tip's definition of ``model slicing''~\cite{tip1994}, which we use to cue the LLM for the slicing task. The second segment is a placeholder to be filled using the textual representation of the Simulink model produced in Step~1 of our approach (see Section~\ref{subsec:conversion}). 

\begin{figure}[t]
\centering
\includegraphics[width=\linewidth]{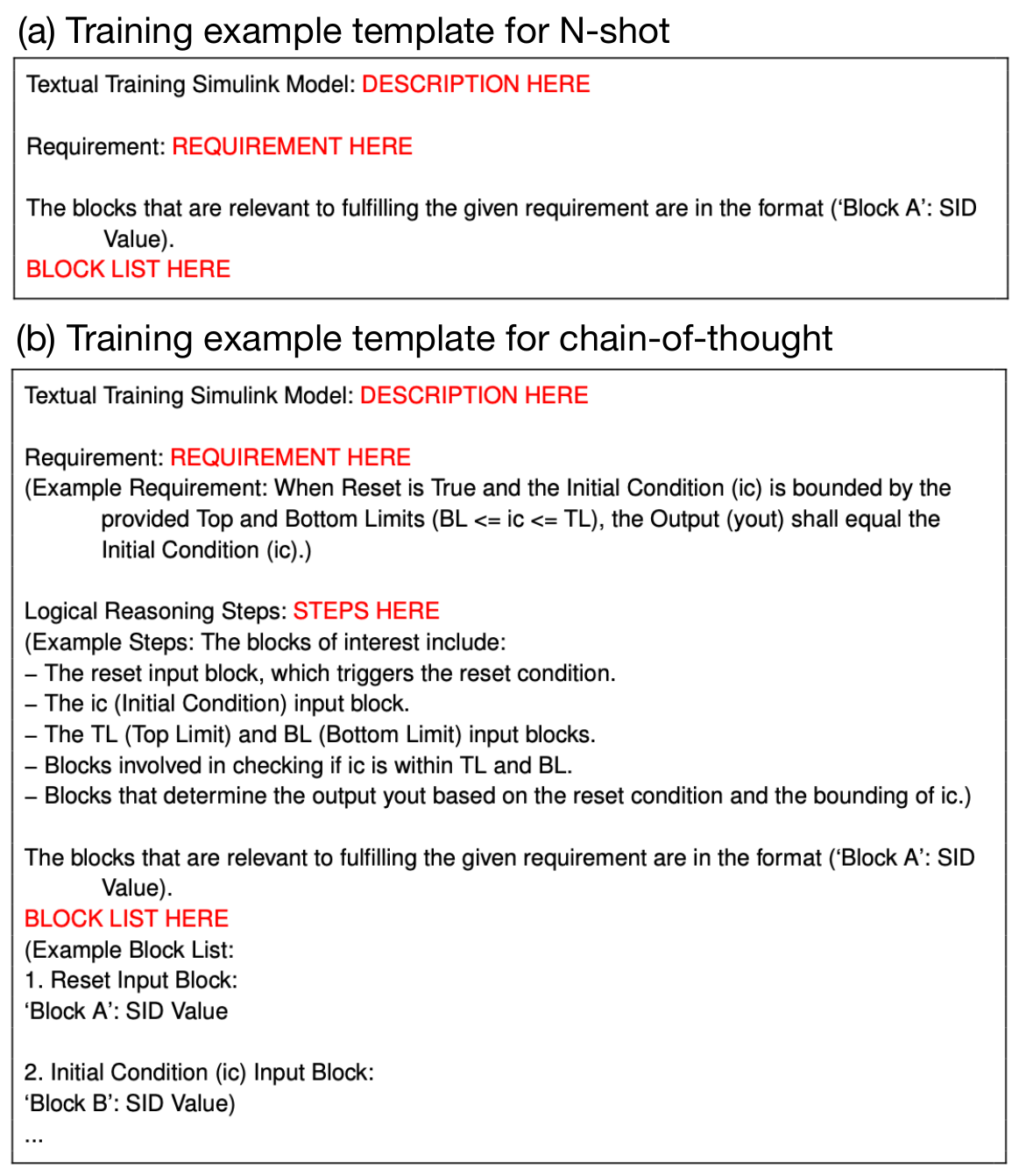}
\caption{Our training-example template for (a) N-shot learning and (b) chain-of-thought learning are comprised of three and four placeholders, respectively. The placeholders, highlighted in red, are instantiated using the input training example(s). The resulting prompt fragments are used to teach the LLM how to create valid model slices.}\label{fig:traning_example}
\end{figure}

The third segment is a placeholder for the  training example(s), if any. 
Note that for zero-shot learning, this segment remains empty.  Figure~\ref{fig:traning_example} shows templates for training examples corresponding to N-shot and chain-of-thought strategies. The templates include red-highlighted placeholders that need to be filled out using a requirements-based slices of the training Simulink models. N-shot learning simply outlines some examples illustrating the model-slicing task. The placeholder for N-shot learning includes: (1) a textual representation of a training Simulink model, (2) a requirement of this (training) model, and (3) a list of block SIDs for the model slice computed for said requirement.  For example, Fig.~\ref{fig:traning_example}(a) shows a template for one-shot learning which involves only one training example. Multiple training examples based on different slices of the training Simulink model(s) can be included in this template to enable N-shot (N$>$1) learning.

As for the chain-of-thought strategy prompting strategy, and as illustrated in Fig.~\ref{fig:traning_example}(b), the training example template includes, in addition to the  three placeholders related to N-shot learning, a placeholder detailing the logical reasoning steps. Given that the chain-of-thought template is more involved than N-shot, we demonstrate the logical reasoning steps. We continue our example with requirement $R$ from Section~\ref{sec:motivatingexample} assuming that the Tustin model is used for training\,\footnote{We note that, in our evaluation of Section~\ref{sec:evaluation}, we use a different model for training, while the Tustin model is the one used for computing slices.}: ``When Reset is True and the Initial Condition (\texttt{ic}) is bounded by the provided Top and Bottom Limits (\texttt{BL} $\leq$ \texttt{ic} $\leq$ \texttt{TL}), the Output (\texttt{yout}) shall equal the Initial Condition (\texttt{ic})''. The logical reasoning steps analyze $R$ to explain how it produces the list of block IDs required to fulfill the requirement. For our example, we note the slice must contain the following block categories from the training model: \texttt{reset}, \texttt{ic}, \texttt{TL} and \texttt{BL}, blocks checking if \texttt{ic} is within \texttt{TL} and \texttt{BL}, and the output \texttt{yout} block. The blocks in \texttt{BlockList} must be explicitly identified as belonging to one of the above \hbox{specified block categories.}

\begin{figure}[t]
\begin{lstlisting}[escapechar=!,basicstyle=\scriptsize\sffamily]
A model slice consists of the parts of a model that potentially impacts the input parameter values computed at some point of interest. This point of interest is referred to as a slicing criterion and is specified by a location in the model in combination with a subset of the model's variables. You are a requirement engineer working on requirements verification and testing for the following system. 

Textual Simulink Model: !\textcolor{blue}{DESCRIPTION HERE}!

Example: !\textcolor{blue}{N-SHOT/CHAIN-OF-THOUGHT EXAMPLE HERE}!

Parse the provided Simulink model to extract the blocks and corresponding SID values which meets the requirement:

Requirement: !\textcolor{blue}{REQUIREMENT HERE}!

Provide your answer as a list of block ids.
\end{lstlisting}
\caption{Our prompt template; the three placeholders in the template are highlighted in blue. The template is instantiated by replacing the blue text to produce LLM prompts for generating Simulink model slices.}
\label{prompt:example}
\end{figure}

The fourth segment in the prompt template of Fig.~\ref{prompt:example} is an instruction to the LLM to retrieve a list of block SIDs relevant to the input requirement \emph{R}, which is included in the prompt through the subsequent (fifth) segment. The final segment is a command instructing the LLM to produce its response as a list of block IDs. The prompt is generated after all placeholders have been filled in. We feed the prompt to a LLM and issue a warning to the user if the prompt's token count exceeds the LLM's input token limit. The LLM extracts a list of necessary blocks from the textual Simulink model to create a model slice for $R$. We use the identified block list to generate a model slice, as we describe next.

\subsection{Slicing}
This step receives the list of blocks deemed necessary for the slice, as identified by an LLM, and extracts a model slice from the given Simulink model. Let \texttt{BlockList} denote the list of blocks returned by the LLM.
We augment \texttt{BlockList} with additional blocks from the original model if necessary to ensure that the slice generated from \texttt{BlockList} is syntactically sound. In particular, two categories of blocks that are not already in \texttt{BlockList} may be included from the original model to ensure that the resulting slice is compilable: (1) Inport and Outport block types, necessary for providing inputs to the model slice and obtaining its output; and (2) components forming a pair with a block in \texttt{BlockList}, such as Goto and From blocks that must be retained together. We refer to blocks from the original model that are absent from \texttt{BlockList} and yet necessary for making the resulting slice executable as ``edge cases''.

Finally, we build a model slice provided with the \texttt{BlockList} and the edge cases. Note that the blocks in \texttt{BlockList} or the edge cases may have dangling input ports when the blocks providing inputs to them are not included in the slice. To resolve dangling inputs, we add a constant block for each such input and link the input to it. In an attempt to maintain behavioural consistency with the original model, we apply the following heuristic to determine the values of the constant blocks: we execute the original model for a randomly selected test input and obtain the value of the input ports in the original model that become dangling in the slice for that test input. The resulting  values are then used for the constant block to fix the dangling inputs in the slice. By doing this, we essentially replace some parts of the original model, not included in the slice, with a constant block. The value of each constant block is set to match the output generated by the part it replaces when executed with a test input. In our evaluation of Section~\ref{sec:evaluation}, we investigate how effective this heuristic is in maintaining consistency with the original model.

For example, suppose we use our approach on the Tustin model and requirement $R$ in Fig.~\ref{fig:tustin} and obtain the following values for \texttt{BlockList}: 89, 90, 96, 132, 140, 93, and 99. Blocks~132 and~140 are of type GoTo, thus requiring the addition of the corresponding From blocks as edge cases. This results in a total of three dangling connections: Switch (block 96), along with the two From blocks (141 and 142). Our approach adds a constant block as input for each of the dangling connections based on an execution of the original Tustin model. The output of this step is a model slice computed based on requirement \emph{R} of the original Simulink model. Figure~\ref{fig:model_slice} presents the model slice derived from the Tustin model in Fig.~\ref{fig:tustin} using our approach.

\section{Evaluation}\label{sec:evaluation}
Our evaluation addresses the following question regarding the use of an LLM-based approach for model slicing.

\textbf{\emph{RQ.} \emph{Are the slices generated by our approach accurate?}}  A crucial question in computing model slices with respect to a requirement is determining whether the slice is accurate. We consider a model slice, computed with respect to a requirement $R$, accurate if it contains sufficient information to determine whether the original model satisfies or violates $R$. In other words, if a slice computed in relation to requirement $R$ satisfies or violates $R$, the original model should also satisfy or violate $R$, respectively. When provided with accurate Simulink model slices, inspection and analysis of individual requirements can focus on their corresponding model slices, eliminating the need to examine an entire model.

To assess the accuracy of the Simulink submodels generated by our approach, we first check if the slice can be compiled. Then, we evaluate the accuracy of the slice using an existing Simulink testing framework~\cite{nejati2019testing}, as discussed in Section~\ref{subsec:accuracy}.  The accuracy of model slices can be impacted by the prompt verbosity-level and the prompting strategy. Thus, we experiment with slices generated based on three different verbosity levels, high, medium and low, and three prompting strategies, zero-shot, N-shot, and chain-of-thought.

\subsection{Accuracy of Simulink Model Slices}
\label{subsec:accuracy}
The goal of our evaluation  is to demonstrate that our generated model slices are accurate, i.e., they preserve enough content from the original model to demonstrate the satisfaction or refutation of the requirements based on which those slices are generated. This requires employing a verification method that can compare the results of assessing requirements on the original model and on model slices. We cannot employ tools based on formal methods, such as model checking~\cite{Clarke:01}, for this purpose, as these tools have known limitations in handling Simulink models. Simulink models often represent dynamic hybrid systems and include continuous and nonlinear mathematical functions, making model checking generally undecidable for Simulink models~\cite{henzinger1998s,alur1995algorithmic,6064535}. In addition, the existing formal verification tools for Simulink are focused on using model checking for generating test cases that achieve some form of structural coverage, e.g., CI/CD coverage, and are not geared towards checking requirements on \hbox{Simulink models~\cite{MatinnejadNBB19}.}

Instead, we use a Simulink model testing framework proposed in the literature~\cite{nejati2019testing,MatinnejadNB17,MatinnejadNBB14} to determine whether a model slice is consistent with the original model in either satisfying or refuting the  requirement of interest. The testing framework is designed to test Simulink models against their requirements. It employs search-based techniques, using meta-heuristic algorithms, to find test inputs likely to expose requirement violations. The framework converts each requirement into a quantitative fitness function, assessing how closely inputs come to violating a requirement. Inputs satisfying the requirement yield a positive fitness value, while those leading to a violation produce a negative or a zero value. We consider a slice generated based on requirement $R$ accurate if assessing the fitness of $R$ yields positive values on both the slice and the original model, or negative/zero values on both. Hence,  requirement $R$ is either met by both the original model and the slice or refuted by both.

For each Simulink model, we consider a test suite for slice accuracy assessment. We then check if, for a given slice computed based on requirement $R$ and for every test case in the test suite, the fitness values pertaining to $R$ consistently match in polarity between the slice and the original model; that is, they are either positive or negative for both the original model and the slice. Note that our slice accuracy assessment approach is partial, as it only ensures the preservation of requirements for test cases within a test suite. To mitigate this partiality, we use Simulink test generation techniques~\cite{MatinnejadNBB19} to maximize the diversity of test cases in the test suite used for slice accuracy assessment. Provided with diverse test cases, we can assess the slice accuracy across different behaviours of the original  model, thereby increasing our confidence in the slices' ability to preserve the behaviours of the \hbox{original model.}

\begin{table*}
\caption{The Simuilnk models used in our experiments: Effector Blender which is used for training and Tustin which is used for testing. The table includes the name of the models, a description of each model, their number of blocks (\#Bk), number of inputs (\#In), number of outputs (\#Out), and number of requirements (\#Reqs).}~\label{tbl:dataset}
\centering
\scalebox{1}{\begin{tabular}{lp{6cm}ccccc}
Name & Description & \#Bk & \#In & \#Out & \#Reqs \\ 
\hline
Effector Blender & A controller that computes the optimal effector configuration for a vehicle. & 95 & 1 & 7 & 3 \\
Tustin & A numeric model that computes integral over time. & 57 & 5 & 10 & 5 \\
\end{tabular}}
\end{table*}

\subsection{Experiment Design}
\label{sec:expdesign}
We use two Simulink models in our experiments: one serves as our training model, while the other is used for evaluation (i.e, computing slices).
Table~\ref{tbl:dataset} presents these two Simulink models: Effector Blender, which encompasses three requirements for our training example template, and the Tustin Integrator, comprising five requirements for  evaluation. In our evaluation, we explore nine distinct  alternatives, combining three levels of model verbosity (high, medium, and low) with three different prompting strategies (chain-of-thought, N-shot, and zero-shot). For each of the nine alternatives, we use our approach (outlined in Section~\ref{sec:appraoch}) to transform our prompt template into a distinct prompt. The prompt is given as input to an LLM to generate a list of blocks essential for fulfilling the prompt's specified requirement, $R$.

In our experiments, as described in Section~\ref{subsec:llm},  we use ChatGPT as the underlying LLM for generating slices. The temperature hyperparameter for ChatGPT controls the balance between predictability and creativity in generated responses. A higher temperature value enables more inventive responses, whereas a lower value ensures consistency. GPT models use a default temperature around 0.7 to achieve a mix of reliability and novelty in  responses. A low temperature restricts the randomness and creativity of an LLM's responses, leading to identical completions across different prompting iterations. As a result, if the LLM generates an incomplete list of blocks required for a slice, it might fail to recognize other relevant blocks in subsequent iterations. Therefore, we use the LLM's default temperature in our experiments to enhance the chances of identifying all relevant elements for the desired slice through multiple iterations. To account for random variation in the LLM's responses, each prompt configuration is repeated three times. We execute 135 iterations of our approach ($9$ configurations x $5$ requirements × $3$ repetitions = $135$). We produce 135 model slices using our slicing method, each corresponding to a distinct run of our approach. In our evaluation, we observed that ChatGPT identifies complementary blocks for slices in the three different repetitions. Consequently, we decided to consider slices not only from the individual repetitions but also from the union of the three repetitions. By aggregating the three iterations for each configuration, we generate an additional 45 slices, totaling 180 slices used in our evaluation ($135 + 45 = 180$). Next, we examine each of these slice to determine whether it is accurate.

As discussed in Section~\ref{subsec:accuracy}, we employ a pre-existing tool~\cite{nejati2019testing} to assess the accuracy of the generated slices. Specifically, we use the tool to generate a test suite with $40$ test cases for the Tustin model to evaluate requirement satisfaction for the generated slices. We then run each of the 180 slices as well as the original Tustin model for these $40$ test cases and compare the resulting fitness values. Our evaluation automatically assesses the behaviour (i.e., fitness value) of a slice in comparison to the original model and identifies a slice as correct if its behaviour matches that of the original model.
It is important to note that a requirement may hold vacuously when a model slice is empty and thereby satisfies requirements that are in the form $p \Rightarrow q$ due to the absence of content~\cite{BeerBER97}. Following the procedure we discuss in Section~\ref{subsec:accuracy}, a slice is accurate if the fitness value has the same polarity in the original model and in the slice, and further, the requirement does not hold vacuously.
Conversely, a requirement is violated if the polarity of fitness values from the original model and the slice are inconsistent, or met vacuously. To calculate the conciseness of a model slice, we count the number of blocks in each slice that non-vacuously satisfies its requirement. The conciseness of a slice gauges its prompt configuration's effectiveness in pinpointing only the blocks essential for satisfying the underlying  requirement. 

\begin{table*}[]
\caption{Accuracy results for 
the slices generated using our approach for the five requirements, \textbf{R1}--\textbf{R5}, of  the Tustin model. The results are presented for nine different prompt configurations, and three different iterations of our approach as well as the union of the three iterations.}\label{tbl:RQ1_table}
\centering
\scalebox{1}{\begin{tabular}{ccccc|cccc|cccc|cccc|cccc}
\multicolumn{1}{l}{} &
  \multicolumn{4}{c}{Requirement 1 (\textbf{R1})} &
  \multicolumn{4}{c}{Requirement 2 (\textbf{R2})} &
  \multicolumn{4}{c}{Requirement 3 (\textbf{R3})} &
  \multicolumn{4}{c}{Requirement 4 (\textbf{R4})} &
  \multicolumn{4}{c}{Requirement 5 (\textbf{R5})} \\
 &
  \textbf{All} &
  \textbf{I1} &
  \textbf{I2} &
  \textbf{I3} &
  \textbf{All} &
  \textbf{I1} &
  \textbf{I2} &
  \textbf{I3} &
  \textbf{All} &
  \textbf{I1} &
  \textbf{I2} &
  \textbf{I3} &
  \textbf{All} &
  \textbf{I1} &
  \textbf{I2} &
  \textbf{I3} &
  \textbf{All} &
  \textbf{I1} &
  \textbf{I2} &
  \textbf{I3} \\
  \hline
\textbf{H-CT} & \checkmark & \checkmark & V & \checkmark & \checkmark & \checkmark & \checkmark & \checkmark & \checkmark & \xmark & \xmark & \xmark & \checkmark & V & \xmark & V & \checkmark & \xmark & \xmark & \xmark \\
\textbf{H-NS} & \checkmark & V & \checkmark & \checkmark & \checkmark & \checkmark & \checkmark & \checkmark & \checkmark & \xmark & \xmark & V & \checkmark & \xmark & V & \xmark & V & V & V & V \\
\textbf{H-ZS} & \checkmark & \checkmark & V & \checkmark & \checkmark & \checkmark & \checkmark & \checkmark & \checkmark & \xmark & \xmark & \xmark & \checkmark & V & \xmark & \xmark & \checkmark & \xmark & \xmark & V \\
\textbf{M-CT} & \checkmark & \checkmark & \checkmark & \checkmark & \checkmark & \checkmark & \checkmark & \checkmark & \checkmark & \xmark & \xmark & V & \checkmark & V & \xmark & \xmark & \checkmark & \xmark & \xmark & V \\
\textbf{M-NS} & \checkmark & V & \checkmark & \checkmark & \checkmark & \checkmark & \checkmark & \checkmark & \checkmark & \xmark & V & \xmark & \checkmark & \xmark & \xmark & V & \checkmark & V & \xmark & V \\
\textbf{M-ZS} & \checkmark & \checkmark & \checkmark & \checkmark & \checkmark & \checkmark & \checkmark & \checkmark & \checkmark & \xmark & \xmark & \xmark & \checkmark & \xmark & \xmark & \xmark & \checkmark & \xmark & \xmark & \xmark \\
\textbf{L-CT} & \checkmark & \checkmark & V & \checkmark & \checkmark & \checkmark & \checkmark & \checkmark & \checkmark & V & \xmark & \xmark & \checkmark & V & V & \xmark & \checkmark & V & \xmark & \xmark \\
\textbf{L-NS} & V & V & V & V & \checkmark & V & \checkmark & \checkmark & \checkmark & \xmark & V & \xmark & \checkmark & V & \xmark & \xmark & V & V & V & V \\
\textbf{L-ZS} & \checkmark & V & V & \checkmark & \checkmark & V & \checkmark & V & \checkmark & \xmark & \xmark & \xmark & \checkmark & \xmark & \xmark & V & \checkmark & V & \xmark & \xmark \\
\\

\multicolumn{1}{l}{} &
  \multicolumn{1}{l}{} &
  \multicolumn{1}{l}{} &
  \multicolumn{1}{l}{} &
  \multicolumn{1}{l}{} &
  \multicolumn{1}{l}{} &
  \multicolumn{1}{l}{} &
  \multicolumn{1}{l}{} &
  \multicolumn{1}{l}{} &
  \multicolumn{1}{l}{} &
  \multicolumn{1}{l}{} &
  \multicolumn{1}{l}{} &
  \multicolumn{1}{l}{} &
  \multicolumn{1}{l}{} &
  \multicolumn{1}{l}{} &
  \multicolumn{1}{l}{} &
  \multicolumn{1}{l}{} &
  \multicolumn{1}{l}{} &
  \multicolumn{1}{l}{} &
  \multicolumn{1}{l}{} &
  \multicolumn{1}{l}{}
\end{tabular}}
\end{table*}
\begin{table*}[!h]
\caption{Size of model slices generated using our approach based on the  prompt configurations from Table~\ref{tbl:RQ1_table}. The table provides the size only for the accurate slices from Table~\ref{tbl:RQ1_table}.}
\label{tbl:RQ2_table}
\centering
\begin{tabular}{ccccc|cccc|c|c|c|c}
\multicolumn{1}{l}{} &
  \multicolumn{4}{c}{\textbf{R1}} &
  \multicolumn{4}{c}{\textbf{R2}} &
  \textbf{R3} &
  \textbf{R4} &
  \textbf{R5} & 
  \textbf{AVG} \\
              & All & I1 & I2 & I3 & All & I1 & I2 & I3 & All & All & All \\
                \hline
\textbf{H-CT} & 14 & 7 & - & 10 & 17 & 10 & 11 & 6 & 9 & 4 & 10 & 10.8 \\
\textbf{H-NS} & 14 & 6 & 13 & 5 & 16 & 11 & 14 & 9 & 8 & 6 & - & 9 \\
\textbf{H-ZS} & 14 & 10 & - & 14 & 21 & 12 & 15 & 11 & 10 & 9 & 9 & 12.6 \\
\textbf{M-CT} & 22 & 7 & 18 & 6 & 10 & 6 & 6 & 10 & 5 & 7 & 15 & 11.8 \\
\textbf{M-NS} & 14 & - & 6 & 7 & 19 & 6 & 11 & 12 & 6 & 5 & 10 & 10.8 \\
\textbf{M-ZS} & 15 & 6 & 13 & 7 & 14 & 12 & 6 & 6 & 9 & 14 & 11 & 12.6\\
\textbf{L-CT} & 12 & 7 & - & 5 & 12 & 7 & 10 & 11 & 9 & 19 & 10 & 12.4 \\
\textbf{L-NS} & - & - & - & - & 11 & - & 4 & 8 & 7 & 8 & - & 7 \\
\textbf{L-ZS} & 9 & - & - & 9 & 10 & - & 10 & - & 9 & 8 & 16 & 10.4 \\
\hline
\textbf{AVG} & 14.3 & 7.2 & 12.5 & 7.9 & 14.4 & 9.1 & 9.7 & 9.1 & 8 & 8.9 & 9.3 & 10.8\\
\multicolumn{1}{l}{} &
  \multicolumn{1}{l}{} &
  \multicolumn{1}{l}{} &
  \multicolumn{1}{l}{} &
  \multicolumn{1}{l}{} &
  \multicolumn{1}{l}{} &
  \multicolumn{1}{l}{} &
  \multicolumn{1}{l}{} &
  \multicolumn{1}{l}{} &
  \multicolumn{1}{l}{} &
  \multicolumn{1}{l}{} &
  \multicolumn{1}{l}{}
\end{tabular}
\end{table*}

\subsection{Results}
Table \ref{tbl:RQ1_table} shows the accuracy results for the slices generated from the Tustin model. Each row of the table  represents a unique configuration, with `H' denoting high verbosity, `M' for medium, and `L' for low verbosity of the textual Simulink model. The suffixes `CT', `NS', and `ZS' correspond to chain-of-thought, N-shot, and zero-shot  strategies, respectively. Each column corresponds to a requirement $R$, i.e., \textbf{R1-R5}, of the Tustin model. The results for the configurations are presented for three individual runs (indicated by \textbf{I1} to \textbf{I3}) as well as the union of the  three iterations (marked as \textbf{All}). Each cell of Table~\ref{tbl:RQ1_table} is marked by \checkmark, indicating the slice and the original model share the same fitness polarity, by \xmark, indicating the polarity of the slice's fitness differs from that of the original model, or by V, indicating that the  slice vacuously satisfies its requirement. 
Table~\ref{tbl:RQ2_table} follows the row and column format of Table~\ref{tbl:RQ1_table} to present the size of each model slice from Table~\ref{tbl:RQ1_table}, provided the slice is accurate, i.e., the behaviour of the slice matches the original model. Cells corresponding to inaccurate model slices are marked with a dash.

We note that all the Tustin model slices in Table \ref{tbl:RQ1_table} can be compiled and executed. Overall, 83 slices out of  the total of 180 generated slices are accurate.  Table \ref{tbl:RQ1_table} shows that slices created by taking the union of iterations -- accounting for LLM response variability -- result in more accurate slices compared to those generated from a single iteration. Union slices are inaccurate in only three out of the 45 cases, occurring only when slices from all three iterations are also inaccurate.

As Table~\ref{tbl:RQ2_table} shows, noting that Tustin has 57 blocks, our approach, on average, reduces the search space for requirement satisfaction to one fifth of the original size \hbox{($57$ / $10.8$ = $5.3$)}.  Slices from low-verbosity configurations are smaller on average for \textbf{R1} and \textbf{R2}, with nine and ten blocks for \emph{L-ZS} compared to an average of 14.3 and 14.4 blocks for the union slices, respectively. However, low-verbosity configurations have the largest model slices for \textbf{R4} and \textbf{R5}, with 19 blocks for \emph{L-CT} and 16 blocks for \emph{L-ZS}, respectively, compared to the overall average slice sizes of 8.9 and 9.3 for \textbf{R4} and \textbf{R5}. Low-verbosity configurations fail to consistently yield concise or accurate slices, evidenced by the highest number of vacuous slices -- 22 slices for the five requirements -- contributing to a total of 38 inaccurate slices out of  a total of 60 low-verbosity configuration slices. This indicates that a highly abstracted (i.e., low verbosity) model lacks the necessary details for an LLM to identify \hbox{relevant blocks of a requirement.} 

Slices generated with a high-level of  verbosity yield a larger number of accurate slices for requirements \textbf{R1} and \textbf{R2}, compared to those from low verbosity configurations. High verbosity configurations only produce three inaccurate, vacuously held slices for \textbf{R1}, with no inaccurate slices identified for \textbf{R2}. Prompts using a high-level of model verbosity do result in the second-highest total of inaccurate slices, totaling to 30 out of 60 slices for the five requirements. High-verbosity configurations incorporate visual rendering details in the textual description of the model, potentially causing the LLM to generate hallucinations~\cite{chen2023hallucination}. Hallucinations can occur when prompts include a substantial amount of data, not all of which is important to the task at hand. When faced with extensive or dense inputs, the LLM may struggle to process all details for pinpointing relevant blocks, leading to the generation of incorrect responses. The medium-verbosity configurations performed the best among the three verbosity levels, resulting in a total of only 28 inaccurate slices out of 60 slices. Hence, our results suggest that using a medium-level of  verbosity for converting a Simulink model into a textual representation yields the most accurate slices.

N-shot configurations result in more concise slices than those generated by chain-of-thought or zero-shot  strategies at the same verbosity level. However, N-shot configurations produce the highest rate of inaccurate slices among the three prompting strategies -- a trade-off we deem unacceptable. Therefore, the prompting choice is between medium-verbosity, chain-of-thought (\emph{M-CT}) and medium-verbosity, zero-shot (\emph{M-ZS}). Our results suggest that \emph{M-CT} and \emph{M-ZS} yield the best performance across the nine prompting configurations. Both \emph{M-CT} and \emph{M-ZS} produced a total of nine inaccurate slices out of a total of 20 slices for the five requirements, as shown in Table~\ref{tbl:RQ1_table}. \emph{M-CT} yields more concise model slices compared to \emph{M-ZS}, with average slice size of 11.8 and 12.6 respectively. A disadvantage of \emph{M-CT} is that it leads to a higher prompt-token count compared to \emph{M-ZS}, due to its need for additional tokens to accommodate a training example. This configuration might not be suitable for large Simulink models due to the increased prompt size, potentially causing the prompt to exceed an LLM's token limit, unlike the equivalent prompt from \emph{M-ZS}.

\vspace*{.5em}
\begin{mdframed}[style=MyFrame]
\emph{The answer to {\bf RQ} is: The medium-verbosity, chain-of-thought \emph{M-CT} and medium-verbosity, zero-shot \emph{M-ZS} prompt configurations result in the largest number of accurate model slices produced by our  approach. Using a medium-level of model verbosity when converting a graphical Simulink model into a textual format ensures a model description with sufficient detail for the LLM to accurately trace the blocks relevant to a requirement, while omitting  visual rendering information that could cause hallucinations in the LLM's response.}
\end{mdframed}
\vspace*{.5em}

\subsection{Limitations and Validity Considerations} \label{sec:threats}
\textbf{Limitations} 
Our preliminary evaluation was conducted using only one training and one testing Simulink model. While these models were selected from an industrial and widely used benchmark for CPS Simulink models~\cite{mavridou2020lockheed}, we emphasize the importance of broadening the scope of evaluation by considering more Simulink models in future experiments.

\textbf{Validity Considerations} The most pertinent validity considerations for our evaluation are internal, external and conclusion validity. 

With regard to \textit{internal validity}, we note that we repeat the experiments for each prompting configuration three times to account for variability in the LLM's responses. To further mitigate random variation, we aggregate blocks from the repetitions  and create additional slices that contain all blocks deemed relevant for each prompting configuration. Our evaluation determines a slice's accuracy by checking whether its fitness value's polarity matches that of the original model. This eliminates the need for any human-provided annotations. The fitness functions used for this evaluation are originate from the existing literature~\cite{nejati2019testing}, which helps reduce bias. It is well-known that LLMs can hallucinate, which means they sometimes produce responses that are inaccurate or inconsistent with reality, even if they seem plausible. In our evaluation, the fitness function helps ensure that the adequacy of a generated slice is not impacted by hallucination. LLMs may nonetheless include in their response  elements that do not belong in the slice. To address this, we explicitly measure conciseness in our evaluation, thus mitigating hidden effects of LLM hallucination.

With regard to \textit{external validity}, we acknowledge that our experiments use a single LLM. The use of different LLMs could affect the accuracy of identifying Simulink blocks within a model slice. More experimentation is required to understand how the choice of LLM impacts the quality of model slicing.

With regard to \textit{conclusion validity}, we note that we cannot conclusively determine whether a slice satisfies a requirement. Our evaluation uses a test suite to assess the behaviour of a slice in comparison to the original model. We do not exhaustively verify the correctness of a slice in preserving its requirement. To mitigate this validity threat, we used the input and output test diversity  criteria for Simulink models~\cite{MatinnejadNBB19} to generate a test suite exercising diverse behaviours of the Simulink model used in our experiments.
\section{Conclusion}\label{sec:conclusion}
In this paper, we proposed an approach based on large language models (LLMs) for requirements-driven slicing of Simulink models. Our approach introduces a prompt template for model slicing, which is systematically instantiated into a specific prompt using the textual representation of a given Simulink model and a natural-language requirement, alongside any training examples. We feed the prompt to an LLM to identify a subset of the blocks from the input model necessary to fulfill the given requirement. Subsequently, we construct a sound slice that incorporates the blocks identified by the LLM. Our preliminary results suggest that prompts developed using Simulink textual representations, which omit visual rendering but retain block content information, combined with chain-of-thought or zero-shot prompting strategies, are the most effective for generating accurate model slices. For future work, we plan to broaden the scope of our evaluation by considering a wider variety of Simulink models and conducting comparisons with baseline methods.

\section*{Acknowledgements}
This work was supported by NSERC of Canada under the Discovery and Discovery Accelerator Programs.

\bibliographystyle{IEEEtran}
\bibliography{ref}
\end{document}